\documentclass[final,3p,times]{elsarticle}

\usepackage{color}
\usepackage{comment}
\usepackage{amsmath}
\usepackage{amsthm}
\usepackage{placeins}

\theoremstyle{plain}

\theoremstyle{definition}
  
\theoremstyle{remark}

\numberwithin{equation}{section}
\numberwithin{lemma}{section}

\journal{Partial Differential Equations in Applied Mathematics}

\begin{document}
  \begin{frontmatter}

    \title{The Dissipative Effect of Caputo--Time-Fractional Derivatives and its Implications for the Solutions of Nonlinear Wave Equations}

    \author[UP,DYSU]{Tassos Bountis}
      \ead{tassosbountis@gmail.com}

    \author[URJC]{Julia Cantis\'an}
       \ead{julia.cantisan@urjc.es }

    \author[US,IMUS]{Jes\'us Cuevas--Maraver}
    \ead{jcuevas@us.es}

    \author[TU,UAA]{J. E. Mac\'{\i}as-D\'{\i}az}
      \ead{jorgmd@tlu.ee, jemacias@correo.uaa.mx}

    \author[UMASS]{Panayotis G. Kevrekidis}
    \ead{kevrekid@umass.edu}

    \address[UP]{Department of Mathematics, University of Patras, 26500 Patras, Greece}
    \address[DYSU]{Center for Integrable Systems, P. G. Demidov Yaroslavl State University, {150003} Yaroslavl, Russia}
    \address[URJC]{Nonlinear Dynamics, Chaos and Complex Systems Group, Departamento de F\'{\i}sica, Universidad Rey Juan Carlos,
    C/ Tulip\'an s/n, 28933 M\'ostoles, Madrid, Spain}
    \address[US]{Grupo de F\'{i}sica No Lineal, Departamento de F\'{i}sica Aplicada I, Universidad de Sevilla. Escuela Polit\'{e}cnica Superior, C/ Virgen de \'{A}frica, 7, 41011-Sevilla, Spain}
    \address[IMUS]{Instituto de Matem\'{a}ticas de la Universidad de Sevilla (IMUS). Edificio Celestino Mutis. Avda. Reina Mercedes s/n, 41012-Sevilla, Spain}

    \address[TU]{Department of Mathematics and Didactics of Mathematics, School of Digital Technologies, Tallinn University, \\ Narva Rd. 25, 10120 Tallinn, Estonia}
    \address[UAA]{Departamento de Matem\'{a}ticas y F\'{\i}sica, Universidad Aut\'{o}noma de Aguascalientes, \\ Avenida Universidad 940, Ciudad Universitaria, Aguascalientes, Ags. 20131, Mexico}

    \address[UMASS]{Department of Mathematics and Statistics, University of Massachusetts Amherst, Amherst, MA 01003-4515, USA}

    \begin{abstract}
      In honor of the great Russian mathematician A. N. Kolmogorov, we would like to draw attention in the present paper to a curious mathematical observation concerning fractional differential equations describing physical systems, whose time evolution for integer derivatives has a time-honored  conservative form. This observation, although known to the general mathematical community \cite {achar2001dynamics, stanislavsky2004fractional, diethelm2010analysis, chung2014fractional, olivar2017fractional, baleanu2020fractional}, has not, in our view,
      been satisfactorily addressed. More specifically,
      we follow the recent exploration of Caputo--Riesz time-space-fractional nonlinear wave equation of~\cite{maciasbountis2022}, in which two of the present authors introduced an energy-type functional and proposed a finite-difference scheme to approximate the solutions of the continuous model. The relevant Klein--Gordon equation considered here has the form:
      \begin{equation}
        \frac {\partial ^\beta \phi (x , t)} {\partial t ^\beta}  - \Delta ^\alpha \phi (x , t) + F ^\prime (\phi (x , t)) = 0, \quad \forall (x , t) \in (-\infty,\infty)
      \end{equation}
      where we explore the sine-Gordon nonlinearity $F(\phi)=1-\cos(\phi)$ with smooth initial data.
      For $\alpha=\beta=2$, we naturally
      retrieve the exact, analytical form of breather waves expected from
      the literature. Focusing on the Caputo temporal derivative
      variation within $1< \beta < 2$ values 
      for $\alpha=2$, however, we observe artificial dissipative effects, which lead
      to complete breather disappearance,
      over a time scale 
      depending on the value of $\beta$.
      We compare such findings to single 
      degree-of-freedom
      linear
      and nonlinear oscillators in the presence
      of Caputo temporal derivatives and also
      consider anti-damping mechanisms to counter
      the relevant effect.
      These findings also motivate some 
      interesting directions for further study, e.g., regarding the consideration of topological solitary waves, such as
      kinks/antikinks and their dynamical
      evolution in this model.
    \end{abstract}

  \end{frontmatter}

\section{Introduction\label{S:Introduction}}

The exploration  of mathematical models involving fractional derivatives has
been progressively expanding and forming  a fruitful area of research
leading to a very substantial volume of publications in many scientific disciplines. To cite a few among the countless relevant examples, in~the epidemiological sciences, relevant  models have been employed to estimate the spreading of computer viruses \cite {singh2018fractional, azam2020numerical}, the~propagation of measles in human populations \cite {qureshi2020real}, as well as
the~transmission dynamics of varicella zoster virus \cite{qureshi2019transmission}. Among numerous other areas,  such applications arise also in
biology (e.g., the~investigation of three-species systems
\cite{ghanbari2020application}), economics (e.g., the~mathematical modeling of Chinese economic growth \cite{ming2019application}) and  nonlinear optics~\cite{malomed}. The relevant progress has been
summarized by now in a number of reviews and
books~\cite{podlubny1999fractional,Samko}, including more specialized
volumes, such as one on nonlinear dispersive waves (see, e.g.,~\cite{cuevas}), which is the topic
of the present work.

Within the
theory of fractional calculus itself,
a wide range of
fractional operators has been proposed \cite {yavuz2020comparing, saad2018new},
leading to the emergence of corresponding analytical results~\cite {ortigueira2021two, ortigueira2021bilateral} in the mathematical literature. However,
a central question continues to focus on the significance and applicability
of such mathematical objects
in real-life physical, chemical or/and biological applications, a theme that is still
very much under investigation. So far, one can argue that Riesz fractional operators have been adequately justified for their use in physical problems \cite {muslih2010riesz} and are commonly used to replace ordinary spatial derivatives. They have been proposed as a valid approach to the continuum limit of systems of particles with long-range interactions~\cite{tarasov2006fractional}
(see also relevant chapters of~\cite{cuevas}, such as the one on the long-range variant of
the sine-Gordon model we will consider below). Additional applications
include, but are not limited to,  the study of long-range interactions on the dynamics and statistics of 1D Hamiltonian lattices with on-site potentials \cite {christodoulidi2018effect} and the phenomenon of supratransmission \cite {macias2018supratransmission, macias2021nonlinear}.

In addition to the Riesz fractional operator, the one due to Caputo  seems to be of great interest to  several areas of science and engineering \cite {jiang2012analytical, chen2013superlinearly, shen2011numerical}. In~particular, these operators have appeared in the extension of well-known parabolic or hyperbolic equations of mathematical physics, including fractional forms of the sine-Gordon equation~\cite {shen2021periodic, bernard1990fractional}, the~nonlinear Klein--Gordon equation \cite {altybay2021fractional}, the~nonlinear Schr\"{o}dinger equation \cite {cuevas,laskin2002fractional}, the fractional KdV equation~\cite{cuevas,INC2020103326}
and~the Klein--Gordon{--}Zakharov system \cite {macias2020existence}. It is worth mentioning that all of these models obey conservation laws which resemble a form of mass or energy conservation of the equivalent systems in the integer-order~case \cite {macias2009numerical}.
It is also worth noting that, from a numerical perspective, a diverse array of
methods have employed different tools, including finite differences \cite {macias2018solution}, finite elements~\cite{wang2021mixed}, spectral techniques \cite {wang2021two}, among~other approaches \cite {pandit2020numerical, mittal2019numerical, mittal2018quasilinearized, shukla2018numerical}.

In a recent work~\cite{maciasbountis2022}, two of the present authors  considered a fractional extension
of the nonlinear Klein--Gordon equation, with Caputo temporal derivatives and Riesz spatial
derivatives and showed that this model satisfies a conservation law in differential form. A
discretization was subsequently introduced that satisfies the same law in the discrete domain and
which leads to a finite-difference model that employs the $L_1$ scheme to approximate the Caputo
temporal derivatives \cite {jin2016analysis}, and~fractional-order centered differences to estimate the
Riesz fractional derivatives in space \cite {ortigueira2006fractional}.
Subsequently,  this nonlinear Klein--Gordon equation was solved numerically
as a Caputo--Riesz time-space-fractional initial-boundary-value problem with homogeneous
Neumann boundary conditions in the form
\begin{equation}
    \label{Eq:Model}%
    \begin{aligned}
      & \frac {\partial ^\beta \phi (x , t)} {\partial t ^\beta}  - \Delta ^\alpha \phi (x , t) + F ^\prime (\phi (x , t)) = 0, \quad \forall (x , t) \in \Omega _T,\\
      & \text {such that } \left\{ \begin{array}{ll}
        \phi (x , 0) = \psi (x), & \forall x \in \overline {\Omega}, \\
        \dfrac {\partial \phi (x , 0)} {\partial t} = \chi (x), & \forall x \in \overline {\Omega}, \\
        \phi (x , t) = 0, & \forall (x , t) \in \partial \Omega \times [0 , T].
      \end{array} \right.
    \end{aligned}
  \end{equation}
with $F(\phi)=1-\cos(u)$
and various values for the $\alpha$ and $\beta$ orders of the Caputo and Riesz
derivatives respectively, using as initial condition the exact breather solution of the $\alpha=\beta=2$ equation \cite {ablowitz1973method} for frequency values $0< \omega < 1$:
\begin{equation}
      \label{Eq:Exact}%
      \phi_\mathrm{br} (x , t) = 4 \arctan \left( \frac {\sqrt {1 - \omega ^2} \cos (\omega t)} {\omega \cosh (\sqrt {1 - \omega ^2} x)} \right), \quad \forall (x , t) \in \mathbb {R} \times [0 , \infty).
    \end{equation}
When the orders of the fractional Caputo and Riesz derivatives are of the classical integer type $\alpha=2$, $\beta = 2$, one naturally gets the expected form of the oscillating breather shown in Fig. \ref{fig:spatial1} below.
The scope of these simulations was to examine the potential persistence of the relevant waveform in the
context of the fractional model.


Our purpose in the present numerical study is to offer further systematic insight on the breather
dynamics in the fractional (in time, via a Caputo derivative) sine-Gordon model. More specifically, we seek to analyze
the dependence on the fractional temporal power $\beta$ and how this leads to a remarkable
decay of the breather. We compare our findings to prototypical lower dimensional fractional
models, such as the one degree-of-freedom fractional pendulum or/and the fractional harmonic oscillator,
while at the same time developing suitable diagnostics including the evolution of the breather
center, its transition time from elastic to viscous behavior, as well as the potential role
of a counter-acting local (dashpot) ``pumping'' and how that can affect the dynamics.
In what follows, we present in section 2, the model setup, while in sections 3 and 4 we expand on our numerical results. Finally, in section 5, we summarize our findings, present our conclusions and discuss some potential directions for future study.

\section{Model setup}

As discussed above, our main aim herein is to develop one of the main
features of the model of \cite{maciasbountis2022}, by considering fractionality solely in the time derivative, via a Caputo fractional derivative. In light of that, we will consider the dynamics of breathers in the time-fractional sine-Gordon equation with a ``classical'' (spatial) dispersion, i.e., for
$\alpha=2$. In other words, our dynamical equation reads:

\begin{equation}\label{eq:FSGE}
    \frac {\partial ^\beta \phi (x , t)} {\partial t ^\beta}  - \frac {\partial ^2 \phi (x , t)} {\partial x ^2} + \sin (\phi (x , t)) = 0
\end{equation}

with $\partial^\beta/\partial t^\beta$ representing the Caputo time-derivative. Depending on the value of $\beta$, several regimes can be observed, even for an uncoupled
(one degree-of-freedom) harmonic oscillator: for $0<\beta\leq1$, there is an overdamped motion and no oscillation is observed; for $1<\beta<2$, which is the most interesting regime and the one we will focus from here on, there is a  motion reminiscent of a damped pendulum, namely, for short times one observes oscillatory motion with decaying amplitude that eventually changes to an overdamped behaviour similar to that of a particle in a viscous medium. Finally, for $\beta>2$, an oscillatory motion with growing amplitude is observed, while, of course, for $\beta=2$ the usual oscillatory behavior is recovered.

For $1<\beta\leq2$, the Caputo derivative is defined as

\begin{equation}
    \frac {\partial ^\beta \phi (x , t)} {\partial t ^\beta} = \frac{1}{\Gamma(2-\beta)}\int_0^t (t-s)^{1-\beta} \left[\frac {\partial ^2 \phi (x , t)} {\partial t ^2}\right]_{t=s} \mathrm{d}s
\end{equation}

Notice that the choice of Caputo derivative instead of, e.g., the Riemann-Liouville derivative is motivated by the fact that we know the initial conditions for integer derivatives. In the next section, we study the evolution of the breather solution \eqref{Eq:Exact} for $t=0$ as initial condition $\phi(x,0)=\psi(x)$ for \eqref{eq:FSGE}. 
Moreover, in what follows, we set $\partial_t\psi(x,0)=0$. {Numerical integration is performed by means of the multiterm \texttt{MT\_FPE\_PI12\_PC} algorithm developed in \cite{Garrappa}.}

\section{Numerical results\label{S:Numerical}}

Let us start by considering the time-evolution of a breather for different values of $\beta$ when $\omega=0.8$ is fixed. To this end, we take as initial condition $\psi(x,0)=\phi_\mathrm{br}(x,0)$ in (\ref{Eq:Exact}), i.e.

\begin{equation}
    \phi(x,0)=4\arctan\left(\frac{\sqrt{1-\omega ^2}}{\omega\cosh(\sqrt{1-\omega^2}x)}\right)
\end{equation}

\begin{figure}[h]
    \begin{center}
    \begin{tabular}{cc}
    \includegraphics[clip,width=0.45\textwidth,trim=0cm 0cm 0cm 0cm]{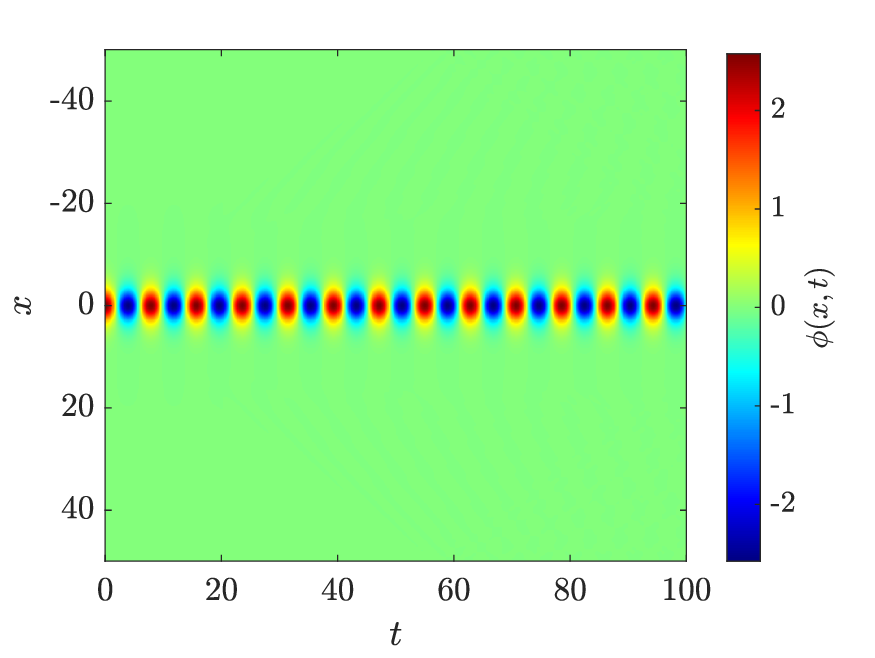} &
    \includegraphics[clip,width=0.45\textwidth,trim=0cm 0cm 0cm 0cm]{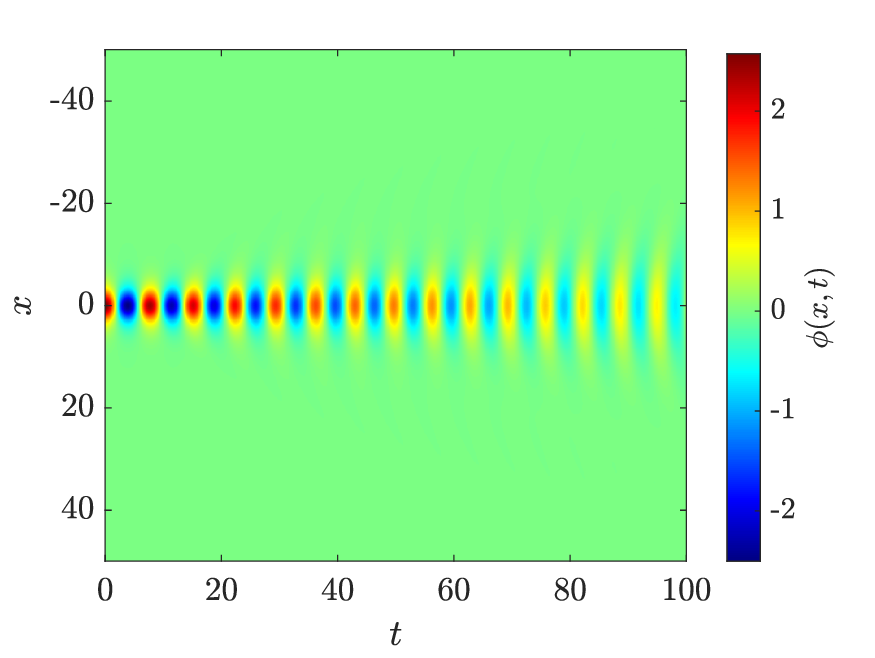} \\
    \includegraphics[clip,width=0.45\textwidth,trim=0cm 0cm 0cm 0cm]{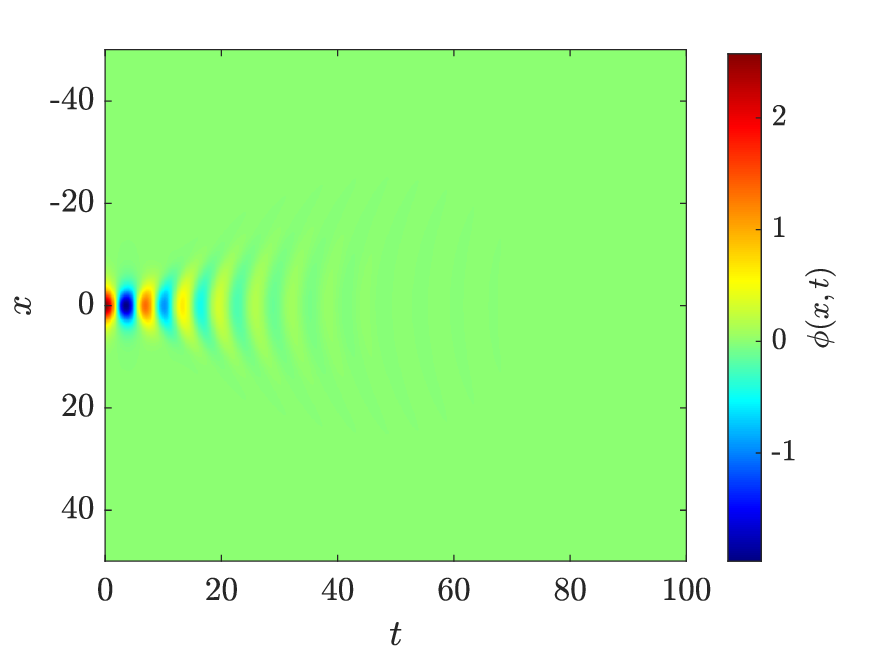} &
    \includegraphics[clip,width=0.45\textwidth,trim=0cm 0cm 0cm 0cm]{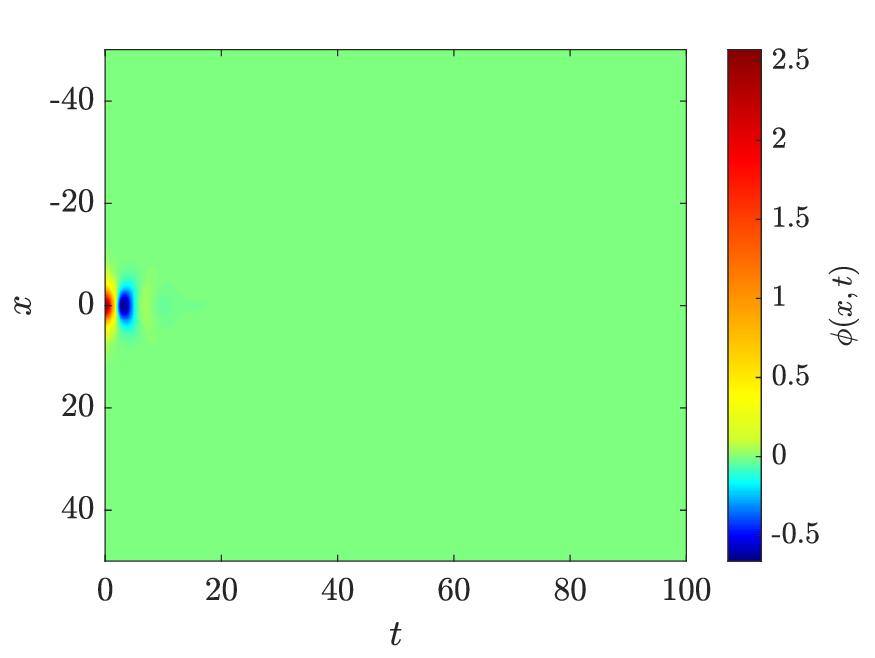} \\
    \end{tabular}
    \end{center}
    \caption{Space-time evolution of the breather for $\omega=0.8$ and $\beta=2$ (top left), $\beta=1.99$ (top right), $\beta=1.9$ (bottom left) and $\beta=1.5$ (bottom right).The rapid decay of the breather
    profile is clearly visible, especially in the latter two panels.}
    \label{fig:spatial1}
\end{figure}

\begin{figure}[h]
    \centering
    \includegraphics[clip,width=0.45\textwidth,trim=0cm 0cm 0cm 0cm]{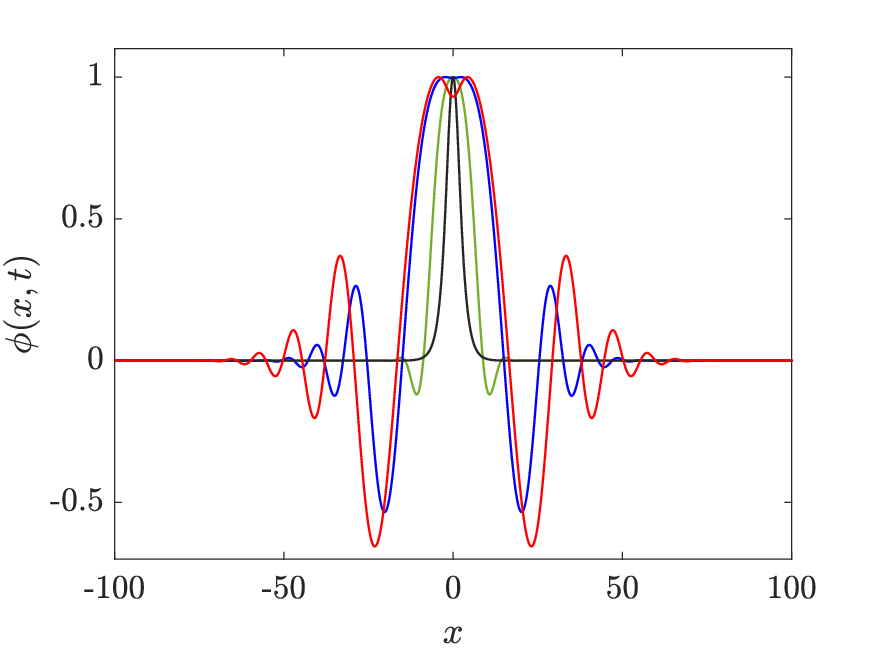}
    \includegraphics[clip,width=0.45\textwidth,trim=0cm 0cm 0cm 0cm]{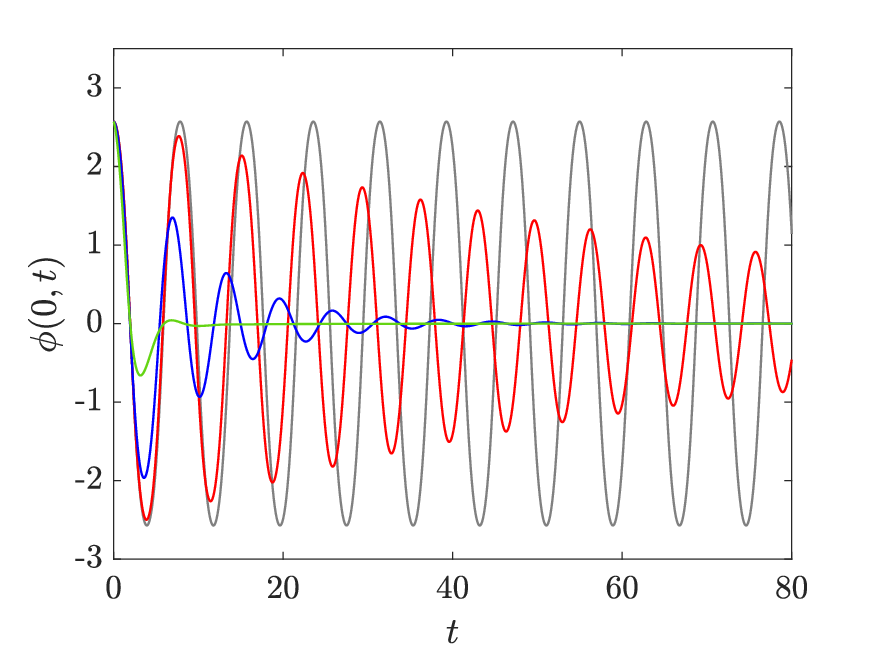}
    \caption{(Left panel) Breather deformation and appearance of secondary "humps" for $\beta=1.9$ and $t=0$ (black), $t=200$ (green), $t=700$ (blue) and $t=950$ (red). The amplitude of the solutions $\phi(x,t)$ decay with time but they are normalized to one in order to compare the evolution.  (Right panel) Amplitude decay of the central node for $\beta=2$ (gray), $\beta=1.99$ (red), $\beta=1.9$ (blue) and $\beta=1.5$ (green).}
    \label{fig:damping1}
\end{figure}

Figure \ref{fig:spatial1} shows the time evolution of such solutions for different values of $\beta$ by means of a density plot; in this case, the displacements (rather than an energy density type quantity) are
plotted. There are two different effects observed in this evolution. On the one hand, the breather deforms
leading to the appearance of secondary ``humps"; 
on the other hand, the breather oscillations are damped.
Indeed, as $\beta$ decreases, the breather survives only for a few oscillations (or, roughly, a single
one as $\beta$ becomes $1.5$, see the bottom right panel of the figure).
The left panel of Fig.~\ref{fig:damping1} shows the deformation of the breather for $\beta=1.9$ by means of several snapshots.
Notice that the maximal profile amplitude has been rescaled to $1$ in this panel,
so as to showcase the
emergence of lateral (secondary) ``humps".

 A more quantitative representation of the damping of the breather oscillations can be seen in the right panel of Fig.~\ref{fig:damping1}, which shows $\phi(0,t)$ for several values of $\beta$. As is evident, fractionality modifies not only the amplitude of the oscillations, but also their frequency $\omega$ which tends to unity, as the oscillation amplitude becomes so small that the linear regime is approached. This change in the frequency depends on $x$, implying loss of coherence of the breather.
 A natural byproduct of the latter is the complete disintegration of the structure observed in Fig.~\ref{fig:spatial1}.

\begin{figure}[h]
    \begin{center}
    \begin{tabular}{cc}
    \includegraphics[clip,width=0.45\textwidth,trim=0cm 0cm 0cm 0cm]{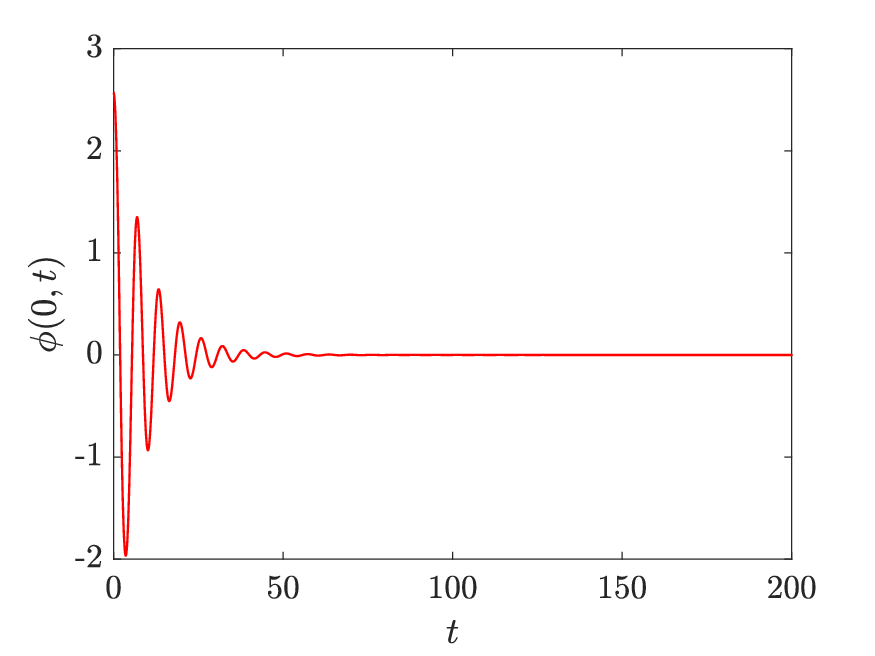} &
    \includegraphics[clip,width=0.45\textwidth,trim=0cm 0cm 0cm 0cm]{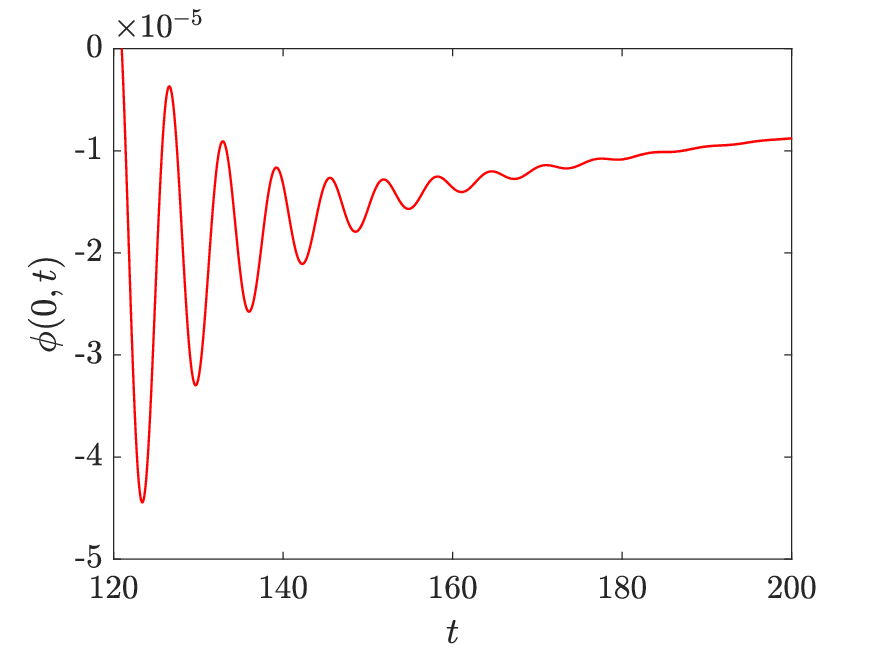} \\
    \includegraphics[clip,width=0.45\textwidth,trim=0cm 0cm 0cm 0cm]{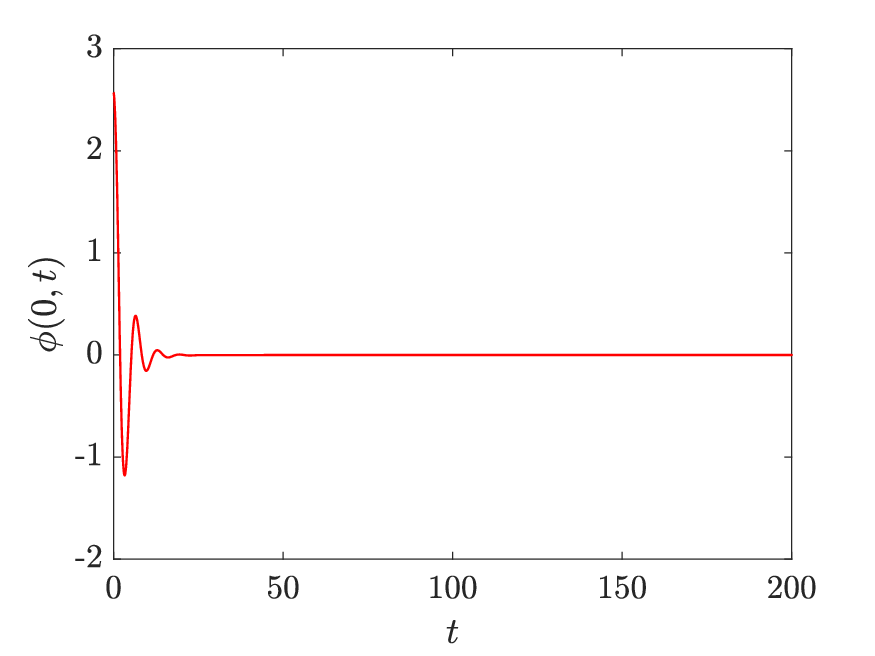} &
    \includegraphics[clip,width=0.45\textwidth,trim=0cm 0cm 0cm 0cm]{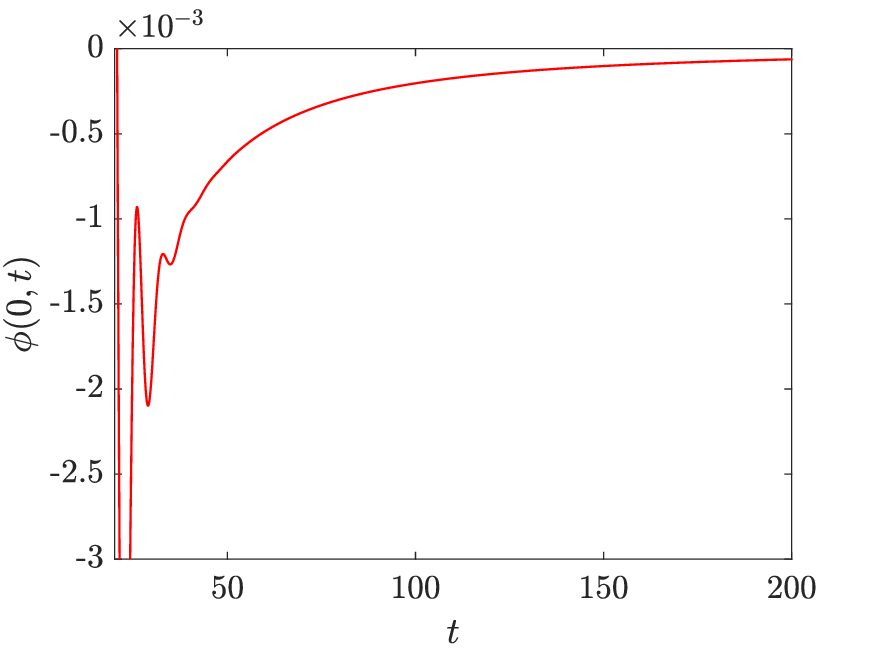} \\
    \end{tabular}
    \end{center}
    \caption{{Evolution of $\phi(0,t)$ for $\omega=0.8$, and $\beta=1.9$ (top panels) and $\beta=1.7$ (bottom panels). The system oscillates around zero for a certain time interval, until $t=t_{s}$, when the system remains compressed and relaxes to zero. Right panels are zooms of left panels for $t>t_{s}$ so that the regime shifts can be observed. For $\beta=1.9$, we find $t_s=120.98$ and for $\beta=1.7$, $t_s=20.67$.}}
    \label{fig:regimes}
\end{figure}

Another interesting feature of the decay process of our fractional model is the existence of two different regimes. If we think of $\phi(x,t)$ as the deformation of a viscoelastic medium, so that it is elongated (compressed) when $\phi>0$ ($\phi<0$), we observe that for short times there are oscillations of the medium (i.e., an elastic regime) around a non-zero value of $\phi$, see left panels of Fig.~\ref{fig:regimes} for two different values of $\beta$. After a time $t_s$, a regime shift can be observed as the medium remains compressed ($\phi<0$ for $t>t_{s}$) and this compression relaxes to zero when $t\rightarrow\infty$, i.e., we are led to a viscous regime. This regime can be visualized in the right panels of Fig.~\ref{fig:regimes}, which are magnifications of the left panels for this regime. This decompression takes the form of oscillations for values of $\beta$ close to $2$ (top right panel), and becomes more
monotonic as one deviates further 
from this limit (bottom right panel).

\begin{figure}[h]
    \centering
    \includegraphics[clip, height=7.5cm,trim=0cm 0cm 0cm 0cm]{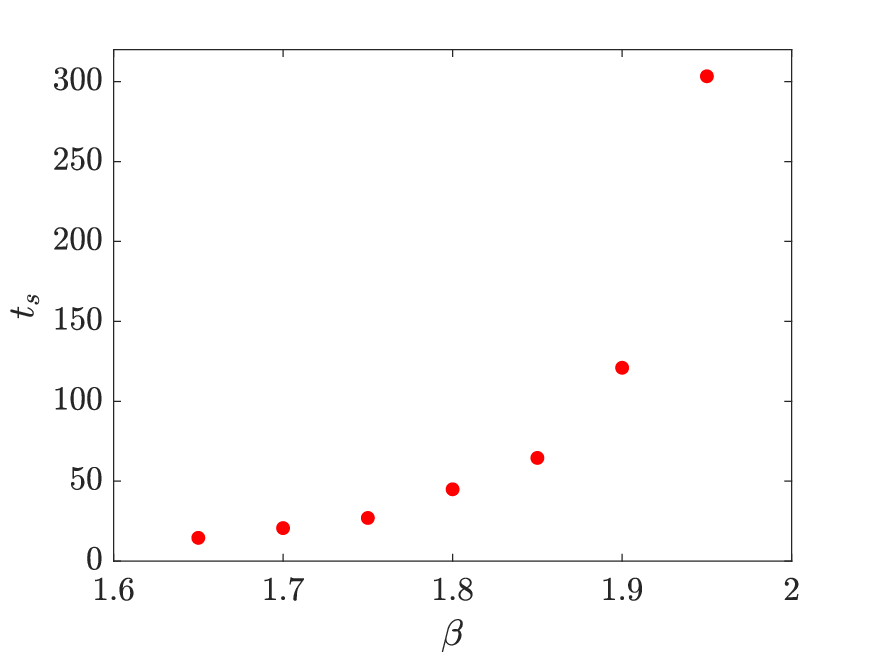}
    \caption{Shift from the elastic to the viscous regime dependence on $\beta$.}
    \label{fig:ts}
\end{figure}

Next, Fig.~\ref{fig:ts} shows the dependence of $t_s$ with respect to $\beta$ for fixed $\omega=0.8$. The value of $t_s$ has been defined as the minimum value of $t$ for which $\phi(0,t)<0$ {whenever} $t>t_s$, {that is, the time at which the medium remains compressed and starts relaxing to zero.
As can be expected, when the integer limit of $\beta=2$ is approached, $t_s$ tends to $\infty$, since in the nonfractional model this distinction of regimes does not exist. For $\beta < 2$, this measure yields a quantitative diagnostic of the deviation from the benchmark scenario.}

We would like to stress that this behavior is different from that of a damped oscillator (or a breather in an elastic medium with losses), where the system oscillates around $0$ for every value of $t$. In fact, we have observed that this ``viscoelastic'' behaviour is not only present for breathers, but also for isolated oscillators either linear or nonlinear.
This strongly suggests that the effect of fractionality bears some qualitative differences in comparison to introducing (regular, linear)
damping to the sine-Gordon equation with $\beta=2$.

\begin{figure}[h]
    \begin{center}
    \begin{tabular}{cc}
    \includegraphics[clip,width=0.45\textwidth,trim=0cm 0cm 0cm 0cm]{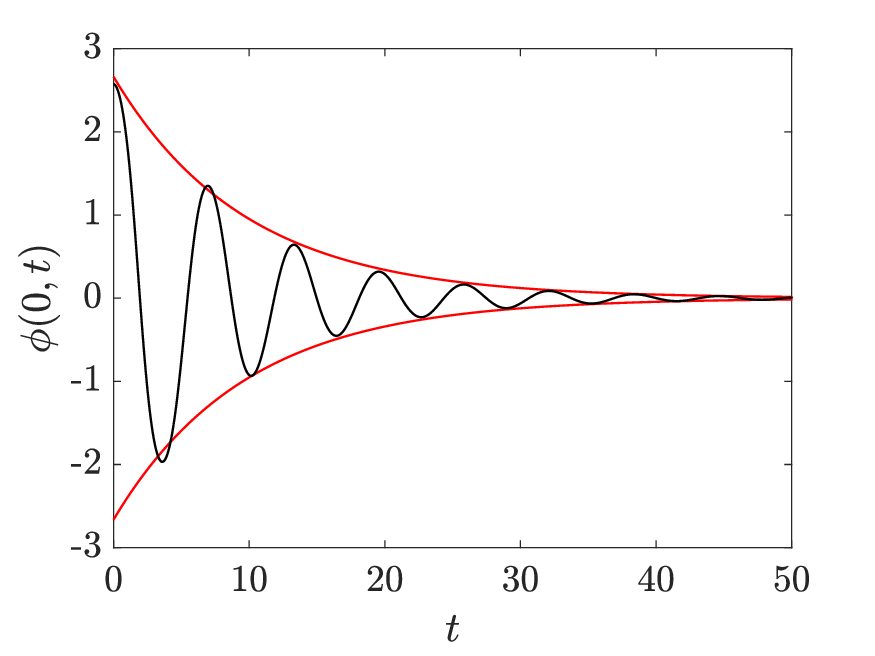} &
    \includegraphics[clip,width=0.45\textwidth,trim=0cm 0cm 0cm 0cm]{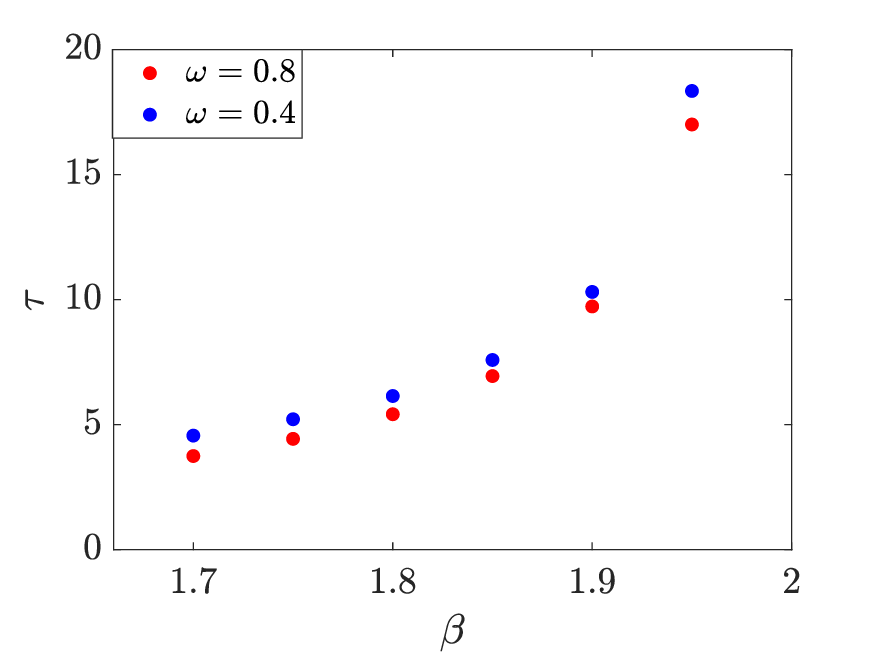} \\
    \multicolumn{2}{c}{\includegraphics[clip,width=0.45\textwidth,trim=0cm 0cm 0cm 0cm]{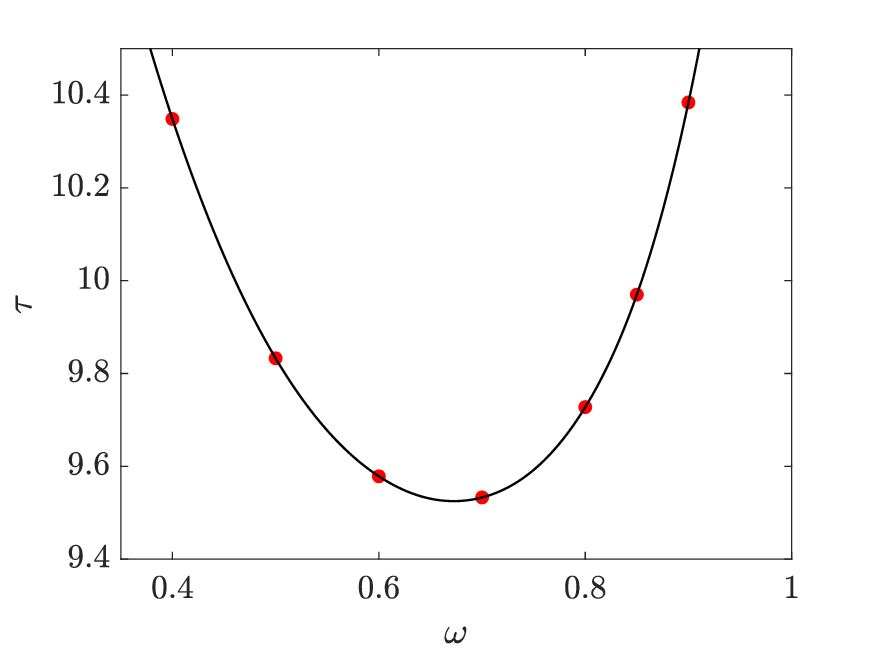}}
    \end{tabular}
    \end{center}
    \caption{(Left panel) Evolution of $\phi(x,0)$ for $\omega=0.8$ and $\beta=1.9$. The envelope of the oscillations is close to an exponential: $ae^{bt}$, with $a=2.6598$, $b=-0.1028$ and $r^{2}=0.997$.  (Right panel) Characteristic decay time $\tau$ for $\omega=0.4$ and $\omega=0.8$ with respect to $\beta$. (Bottom panel) Decay times for different values of the frequency for fixed $\beta=1.9$. The black line is just a fit to a cubic spline interpolation to guide the view.}
    \label{fig:damping2}
\end{figure}

Once we have qualitatively identified the dynamics of breathers, we proceed to analyze in more detail the observed decay in the elastic regime {(i.e., for $t<t_s$)}. In Fig.~\ref{fig:damping2} we fit the extrema of the oscillation amplitude of $\phi(0,t)$ to an exponential, i.e. $\phi(0,t)\sim a e^{bt}$, with $\tau=-1/b$ being the characteristic decay time. This figure also shows the dependence of $\tau$ with respect to the fractionality $\beta$ for two values of the breather frequency $\omega$. One can see that the dependence of $\tau(\beta)$ is qualitatively similar for both values of $\omega$.

Naturally, also $\tau\rightarrow\infty$ when $\beta\rightarrow2$, as one approaches the classical (and Hamiltonian) sine-Gordon equation. If we now  vary the frequency $\omega$ and fix $\beta=1.9$, we obtain a surprisingly non-monotonic behavior concerning the dependence of $\tau$ on $\omega$ (see bottom panel of Fig.~\ref{fig:damping2}). Fractionality seems to be more efficient as a damping ``mechanism'' for certain frequencies of the breather. Specifically, we obtain a minimum of $\tau$ close to $\omega=0.7$. A similar trend is found for other values of $\beta$. Despite the fact that the relevant variation in $\tau$ is small, we do not have a ``structural'' explanation of this effect and attribute it, at least in part, to the particular functional form of the breather.

In order to study if the previous observations related to damping are particular to fractional breathers, we now analyze the breather dynamics in connection to two fractional one-degree-of-freedom systems: a linear
(harmonic oscillator) and a nonlinear one (pendulum). The latter one is of the form:

\begin{equation}\label{eq:pendulum}
    \frac{\mathrm{d}^\beta u (t)}{\mathrm{d} t^\beta}+\sin(u(t)) = 0.
\end{equation}

Our initial conditions for this system are taken as $u(0)=u_0$, $\dot{u}(0)=0$, where the oscillating amplitude $u_0$ is related to the oscillation frequency $\omega$ through

\begin{equation}
    \omega=\frac{\pi}{2K(m)(\sin^2(u_0/2))},
\end{equation}

$K(m)$ being the complete elliptic integral of the first kind with parameter $m$. Thus, $u_{0}$ can be calculated for any desired value of the frequency.

In the small amplitude limit, the fractional pendulum can be described by a fractional harmonic oscillator:

\begin{equation}\label{eq:HO}
    \frac{\mathrm{d}^\beta u_\mathrm{HO} (t)}{\mathrm{d} t^\beta}+\omega^\beta u_\mathrm{HO}(t) = 0.
\end{equation}

In this case, the solution is known and can be expressed as a series expansion of Mittag-Leffler functions $E_\beta(z)$ \cite{Mainardi},

\begin{equation}
    u_\mathrm{HO}(t)=u_0 E_\beta[-(\omega t)^\beta],\qquad E_\beta(z)=\sum_{n=0}^\infty \frac{z^n}{\Gamma(\beta n+1)}.
\end{equation}

Unlike the pendulum, in the  harmonic oscillator, the frequency and the decay of oscillations are not affected by the initial amplitude. Thus, we can fix the initial condition  to ($u_0=1$, $\dot{u}_0=0$), without loss of generality.

\begin{figure}[h]
    \begin{center}
    \begin{tabular}{cc}
    \includegraphics[clip,width=0.45\textwidth,trim=0cm 0cm 0cm 0cm]{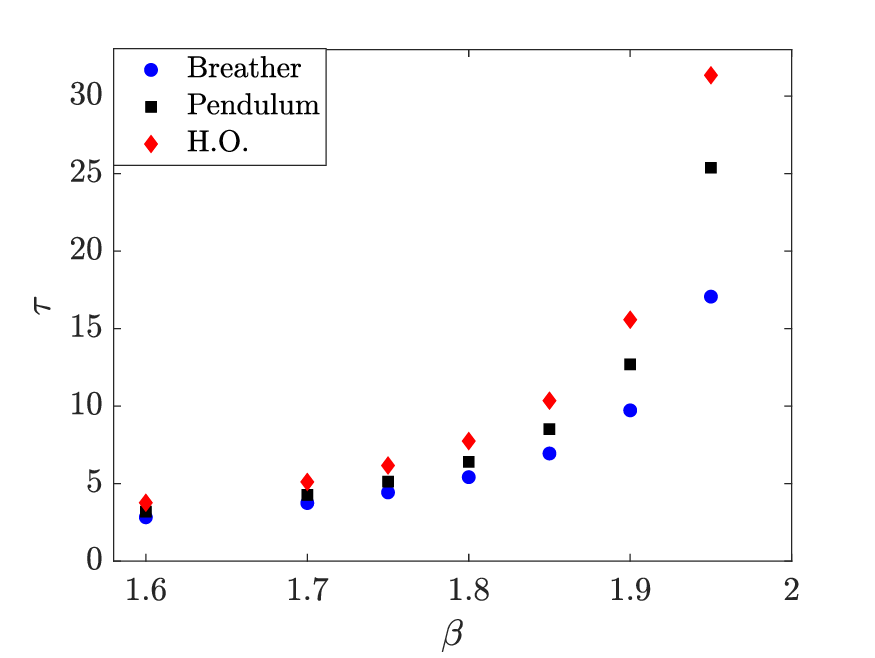} &
    \includegraphics[clip,width=0.45\textwidth,trim=0cm 0cm 0cm 0cm]{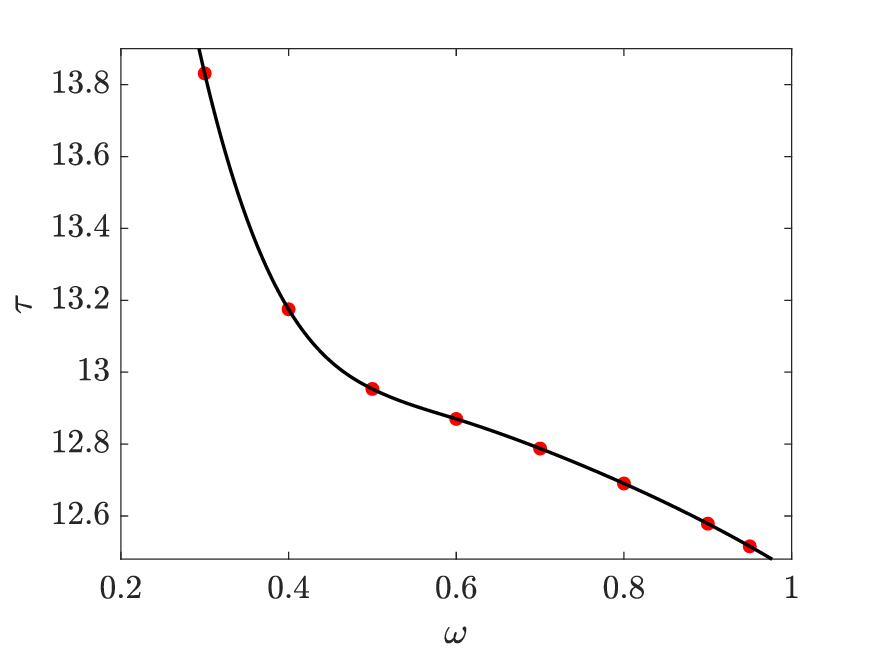} \\
    \end{tabular}
    \end{center}
    \caption{(Left panel) Characteristic decay time $\tau$ with respect to $\beta$ for $\omega=0.8$ in the breather, pendulum and harmonic oscillator cases. Notice that for the HO, $\omega=\omega_0^\beta=0.8$ (Right panel) Decay times for different values of the frequency for fixed $\beta=1.9$ in the pendulum case. The black line is just a fit to a cubic spline interpolation to guide the view.}
    \label{fig:pendulum1}
\end{figure}

Interestingly, both the pendulum and the harmonic oscillator exhibit dynamics similar to the breather regarding  their decay behaviour.  Figure \ref{fig:pendulum1} compares the decay times of the breather, pendulum and harmonic oscillator with frequency $\omega=0.8$ with respect to $\beta$.
It is important to observe that this decay time is shortest for the case of the breather, which suggests that the role of the additional degrees of freedom is to lead the solutions to a faster ``decoherence''. Furthermore, the nonlinear pendulum model is more rapidly decaying than the harmonic oscillator, suggesting that the nonlinearity enhances the dissipative dynamics in this context.
On the right panel of Fig. 6, the decay time with respect to the frequency is shown  for $\beta=1.9$ in the case of the pendulum. One can see that, contrary to the breather case, this dependence is monotonic. This is also the case for the harmonic oscillator. Once again, as indicated above, the variations of the relevant
time scale $\tau$ are rather small, yet there is a clear distinction of the ``collective'' sine-Gordon model in comparison to the single degree of freedom systems for sufficiently large frequencies,
i.e., for $\omega>0.7$ (where the minimum of the quantity $\tau$ occurs for the sine-Gordon case).

\section{The effect of anti-damping terms\label{S:Antidamping}}

Finally, we discuss the possibility of adding an anti-damping term to (\ref{eq:FSGE}) to avoid the effects of fractionality, i.e., the emergence of exponential damping, and secondly,  the appearance of a regime shift that separates the evolution of the system in a viscoelastic and a viscous phase. We do not expect to avoid the latter effect, but counter-acting the exponential damping with such an anti-damping term is of interest in its own right. To this end we study the following system:

\begin{equation}\label{Eq:antidamp}
    \frac {\partial ^\beta \phi (x , t)} {\partial t ^\beta}  - \frac {\partial ^2 \phi (x , t)} {\partial x ^2} + \sin (\phi (x , t)) + \gamma \frac {\partial \phi (x , t)} {\partial t}= 0,
\end{equation}
where the anti-damping corresponds to the last term, with $\gamma<0$.

\begin{figure}[h]
    \begin{center}
\includegraphics[clip,width=0.45\textwidth,trim=0cm 0cm 0cm 0cm]{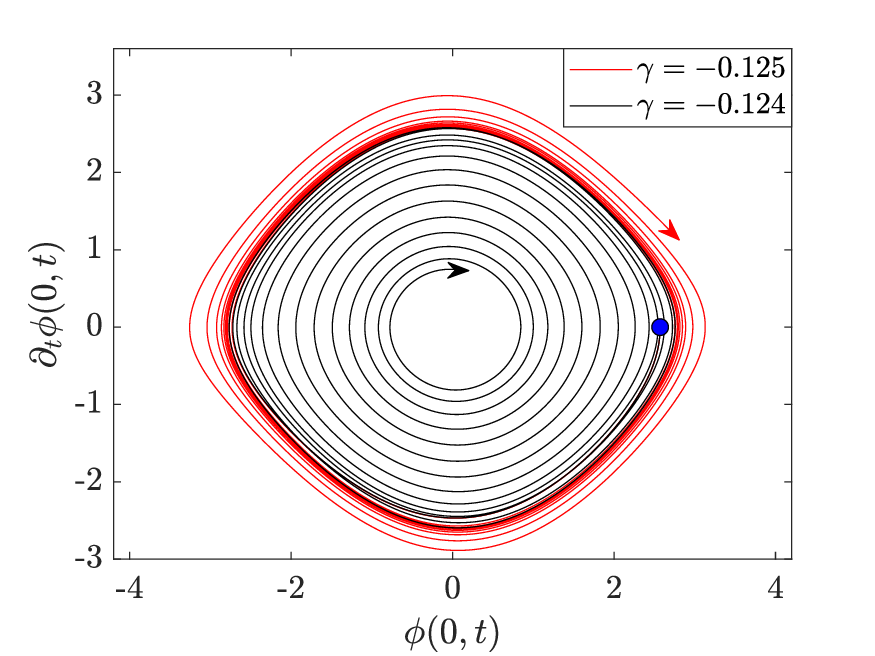}
\includegraphics[clip,width=0.45\textwidth,trim=0cm 0cm 0cm 0cm]{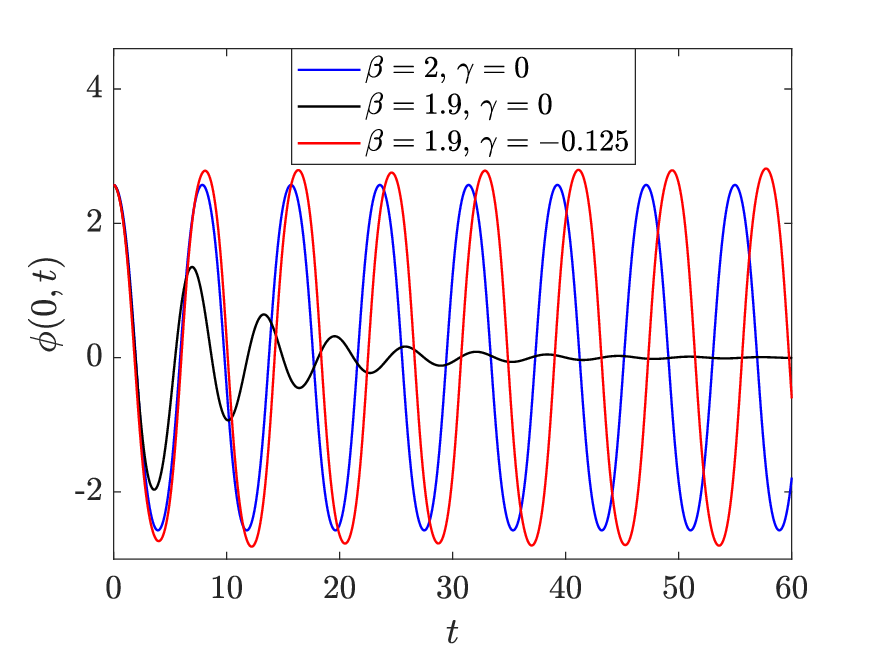} \\
\includegraphics[clip,width=0.45\textwidth,trim=0cm 0cm 0cm 0cm]{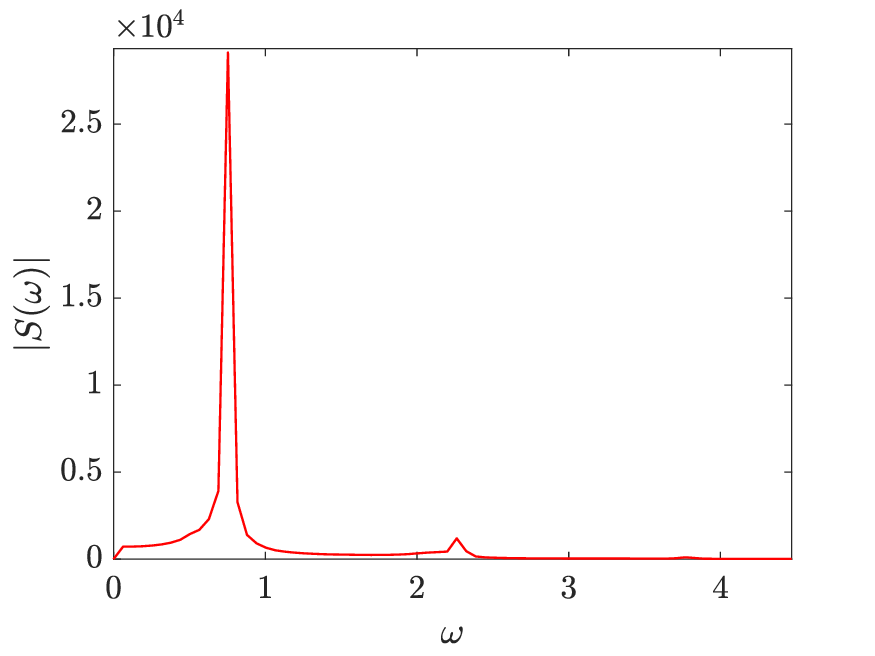} {\includegraphics[clip,width=0.45\textwidth,trim=0cm 0cm 0cm 0cm]{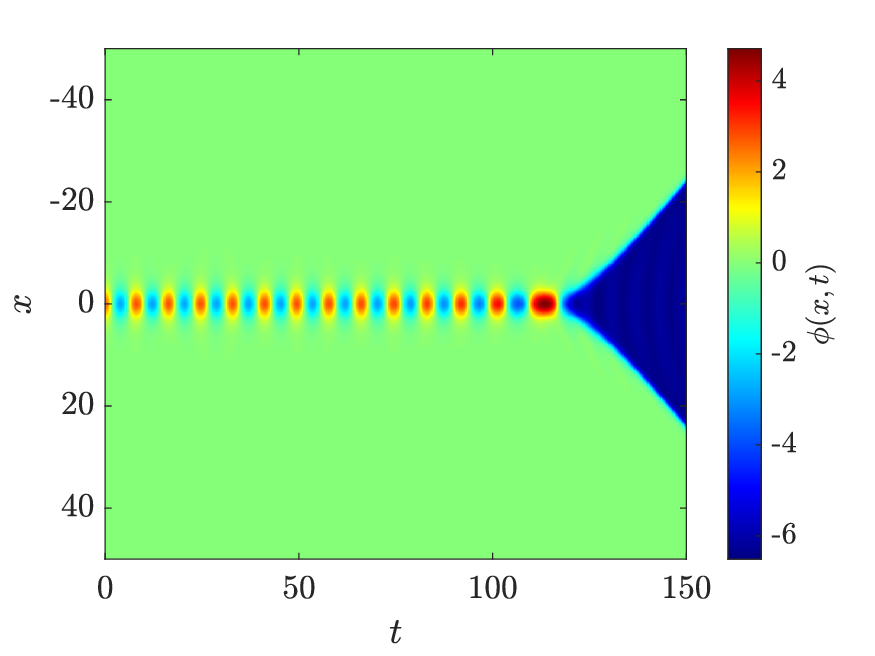}} \\
    \end{center}
    \caption{(Top left panel) Trajectories in phase space for $x=0$ and values of $\gamma$ larger and smaller than $\gamma_{c}$. The blue dot denotes the common initial condition. At the beginning both trajectories follow the same path, but after a while, one is amplified while the other one is damped. (Top right panel) Time evolution for the breather with frequency $\omega=0.8$ without fractionality (blue line), with fractionality (black line) and with fractionality and antidamping (red line). Fractionality induces both damping and a change
      in the frequency. The antidamping may restore the amplitude but not the frequency. {(Bottom left panel) Frequency spectrum for $\phi(0,t)$ with $\beta=1.9$ and $\gamma=-0.125$.} (Bottom right panel) Space-time evolution of the breather for $\beta=1.9$ with $\gamma=-0.125$. For a transient time, oscillations with slightly increasing amplitude are recovered, before the configuration eventually breaks up
      into a pair of moving kinks. In all panels, $\beta=1.9$, $\omega=0.8$.}
    \label{fig:anti1}
\end{figure}

We observe that from a certain value on, the damping induced by the fractionality disappears and oscillations are maintained or amplified.
For $\beta=1.9$ and $\omega=0.8$, this value is $\gamma_{c} \approx -0.1245$. In the top left panel of Fig.~\ref{fig:anti1} we can see trajectories in phase space $(\phi(0,t),\partial_t\phi(0,t))$ for values of $\gamma$ larger and smaller than $\gamma_{c}$.
It is clear that the case below $\gamma_c$ still exhibits damping,
while the other one slightly above counter-balances this effect and leads to some residual, weak
pumping.
However, as shown in the top right panel of Fig.~\ref{fig:anti1}, the antidamping is not able to restore the initial frequency which was modified by the fractionality. In fact, we observe that for the blue line ($\beta=2$, $\gamma=0$), the frequency is $\omega=0.8$. When the fractionality is introduced, which corresponds to the black line ($\beta=1.9$, $\gamma=0$), the frequency is modified to $\omega=1$.
Finally, when we add the antidamping term, which is represented by the red line ($\beta=1.9$, $\gamma=-0.125$), oscillations no longer decay, in fact, they are amplified, but the frequency is changed to
approximately $\omega=0.75$. {The frequency spectrum is depicted in the bottom left panel of Fig.~\ref{fig:anti1}. As it can be seen, there is a small peak at $3\omega$=2.25, which corresponds to one of  the harmonics of the main frequency peak. For other values of $x \neq 0$, close to the central node, we obtain the same frequency spectrum. These facts suggest that a frequency-shifted breather is generated. This needs to be explored in the future by means of fixed point algorithms.}

It is also interesting to notice that for $\gamma>\gamma_{c}$ the dynamics of the system becomes more complex and as the amplitude increases the system may jump to a different attractor. Thus, the oscillations seen in Fig.~\ref{fig:anti1} for $\gamma=-0.125$ are only a transient phenomenon.
At longer times, as is it observed in the bottom right panel, the structure breaks down into a pair of kinks that
subsequently move outward  from $x=0$.
We conclude that fractionality perturbs the system in a way that the antidamping term cannot fully amend. 

\section{Conclusions and future challenges\label{S:Conclusion}}

In a recent paper \cite {maciasbountis2022}, a subset of the present authors
considered a generalized nonlinear wave equation in multiple dimensions which extends well-known
conservative models of mathematical physics, such as the sine-Gordon equation,
from their classical to their fractional form. In that work, fractional temporal
derivatives of the Caputo type and spatial Riesz-fractional derivatives were utilized.
In the same work \cite {maciasbountis2022}, the existence and uniqueness of solutions
was proved for this model using a fixed-point theorem, and it was verified that a consistent discretization of the conserved quantity is also preserved through time.

Our main purpose in the present paper is to emphasize the dissipative nature of the Caputo time-fractional derivative, a feature that was present but not further detailed in the earlier
work of~ \cite {maciasbountis2022}. This effect is present in numerous of the results discussed here for the case of a time-fractional sine-Gordon equation and order of differentiation $\beta < 2$. It is important to note that the sine-Gordon equation remains a conservative model, even when space-fractional derivatives of the Riesz type are considered. Evidently, therefore, it is the presence of the Caputo time-fractional derivatives that turns this system into a dissipative model. 
 Analogous results have also been observed in the literature \cite {achar2001dynamics, stanislavsky2004fractional, diethelm2010analysis, chung2014fractional, olivar2017fractional, baleanu2020fractional} mostly in cases where ordinary (rather than partial) differential equations are studied.
Relevant explicit results are available, e.g., in the case of linear harmonic oscillator
(e.g., in the work of~\cite{diethelm2010analysis}) showcasing the decreasing amplitude
of the relevant model.

Here, we have taken these results further, not only developing spatial and temporal
diagnostics for the evolution of the breather, but also offering comparisons with the single DOF
models, illustrating the role of the nonlinearity and of the coupling to the additional
degrees of freedom (in comparison to a single pendulum) towards decreasing the lifetime of
the pertinent breathing oscillations. Furthermore, we have considered a ``counter-action''
mechanism through an anti-damping, and have clarified which aspects of the evolution we
can hope to ``restore'' (such as, e.g., the amplitude) and which remain modified
(such as, e.g., the frequency).


Thus, it is clear that a deeper analysis of this ``dissipative" phenomenon needs to be carried out in future works concerning the solutions of conservative systems of fractional differential equations, whose derivatives with respect to time are of the Caputo type.
Furthermore, both the present work and that of~ \cite {maciasbountis2022} offer additional
insights for possible future studies. In particular, in the presence of anti-damping herein,
we have observed the breakdown of the breather structure towards a kink-antikink pair.
On the other hand in \cite {maciasbountis2022}, it was observed that the combined
action of the Caputo time derivative and the Riesz spatial derivative (when suitably
deviating from the sine-Gordon limit) can lead to nearly stationary kink-antikink-like
bound states. Both of these results, but also the topological nature of the kink structures,
suggest the particular interest and relevance of the study of such topological states and
their multi-kink generalizations.
Indeed, it is important to recall that,  unlike the breather, the kink
cannot be subject to destruction (although it can be, e.g.,
subject to deceleration and stopping, akin to what happens in discrete variants of the sine-Gordon
model~\cite{PEYRARD198488}). This is a natural next step for our considerations which is
presently under study and which will be reported in future publications.

\FloatBarrier
  \paragraph*{Author contributions}

 Conceptualization, T.B., J.E.M.-D.; formal analysis, J.C., J.C-M. P.G.K.; methodology, J.C., J.C-M., P.G.K.; project administration, T.B., P.G.K.; software, J.C., J.C-M.; supervision, T.B., P.G.K.; validation, T.B., J.C.,J.C-M., J.E.M.-D. P.G.K.; writing---original draft, T.B., J.C., J.C-M., P.G.K.; writing---review and editing, T.B., J.C., J.C-M., J.E.M.-D., P.G.K. All authors have read and agreed to the published version of the~manuscript.
   
   \paragraph*{Funding}

  J.E.M.-D.: The present work reports on a set of final results of the research project ``Conservative methods for fractional hyperbolic systems: analysis and applications'', funded by the National Council for Science and Technology of Mexico (CONACYT) through grant A1-S-45928.
  This material is based upon work supported by the U.S. National Science Foundation under the awards PHY-2110030 and DMS-2204702 (PGK). J.C.-M. acknowledges support from the EU (FEDER program 2014–2020) through MCIN/AEI/10.13039/501100011033 (under the projects PID2020-112620GB-I00 and PID2022-143120OB-I00).
  J.C. acknowledges that this work has been supported by the Spanish State Research Agency (AEI) and the European Regional Development Fund (ERDF, EU) under Project No. PID2019105554GB-I00 (MCIN/AEI/10.13039/ 501100011033).

  \paragraph*{Data availability}

  The data presented in this study are available on request from the authors.

  \paragraph*{Conflict of interest statement}

  The authors declare no potential conflict of interest.

  \bibliographystyle{elsarticle-num}
  \bibliography{Bibfile}

\end{document}